\documentclass[mathleft]{an}
\usepackage{graphicx}
\usepackage{times}
\usepackage{url}             
\begin{document}

\Pagespan{1}{}
\Yearpublication{2006}%
\Yearsubmission{2005}%
\Month{11}%
\Volume{999}%
\Issue{88}%

\title{Update on g-mode research}

\author{R. A. Garc\'\i a\inst{1}\fnmsep\thanks{Corresponding author:
  \email{rgarcia@cea.fr}\newline}
\and  A. Jim\'enez\inst{2}
\and S. Mathur\inst{1}
\and J. Ballot\inst{3}
\and A. Eff-Darwich\inst{2,4}
\and S. J. Jim\'enez-Reyes\inst{2}
\and P. L. Pall\'e\inst{2}
\and J. Provost\inst{5}
\and S. Turck-Chi\`eze\inst{1}
}
\titlerunning{Update on g-mode research}
\authorrunning{Garc\'\i a et al.}
\institute{Laboratoire AIM, CEA/DSM-CNRS - U. Paris Diderot - IRFU/SAp, 91191 Gif-sur-Yvette Cedex, France
\and 
Instituto de Astrof\'\i sica de Canarias (IAC), 38205, La Laguna, Tenerife, Spain
\and Max-Planck-Institut f\"ur Astrophysik, Karl-Schwarzschild-Strasse 1, 85748 Garching, Germany 
\and Departamento de Edafolog\'\i a y Geolog\'\i a, Universidad de La Laguna, Tenerife, Spain
\and D\'epartement Cassiop\'ee, UMR CNRS 6202, Observatoire de la Cote dÕAzur, BP 4229, 06304 Nice Cedex 4, France
}
\received{October 2007}
\accepted{February 2008}
\publonline{2008}


\abstract{
Since the beginning of this century we have attended a blooming of the gravity-mode research thanks to the unprecedented quality of the data available, either from space with SoHO, or from the ground-based networks as BiSON or GONG. From the first upper limit of the gravity-mode amplitudes fixed at 10 mm/s at 200 $\mu$Hz given by Appourchaux et al. (2000), on one hand, a peak was supposed to be a component of the $\ell$=1, $n$=1 mixed mode (Garc\'\i a et al. 2001a, b; Gabriel et al. 2002) and, on the other hand, a couple of patterns --multiplets-- were attributed to gravity modes (Turck-Chi\`eze et al. 2004; Mathur et al. 2007). One of these patterns, found around 220 $\mu$Hz, could be labeled as the $\ell$=2, $n$ =-3 g mode, which is expected to be the one with the highest surface amplitude (Cox and Guzik 2004). Finally, in 2007, Garc\'\i a et al. were able to measure the fingertips of the dipole gravity modes looking for their asymptotic properties. In the present paper we present an update of the recent developments on this subject with special attention to the 220 $\mu$Hz region, the dipole asymptotic properties and the impact of the incoming g-mode observations on the knowledge of the solar structure and rotation profile. }
\maketitle

\section{Introduction}
 In the early 80's, right after the beginning of the helioseismology, the gravity-mode (g modes) searches star\-ted. Several groups looked for both individual modes and the signature of their asymptotic properties (see for example: De\-la\-che and Scherrer (1983); Fr\"ohlich and Delache (1984); Isaak et al. (1984); Pall\'e and Roca-Cort\'es (1988) and the reviews by Hill et al. (1991) and Pall\'e (1991)). Unfortunately, none of these candidates could be confirmed as gravity modes by more recent observations. In the middle of the 90's, two new observational facilities became fully operational in helioseismology: the complete deployment of the ground-based networks (e.g. BISON\footnote{Birmingham Solar Oscillation network (Chaplin et al. 1996)} and GONG\footnote{Global Oscillation Network Group (Harvey et al. 1996)}) and the la\-unch of SoHO\footnote{Solar and Heliospheric Observatory (Domingo et al. 1995)} (with three helioseismic expe\-ri\-ments: GOLF\footnote{Global Oscillations at Low Frequency (Gabriel et al. 1995)}, VIR\-GO\footnote{Variability of solar IRradiance and Gravity Oscillations (Fr\"olich et al. 1995)} and SOI/MDI\footnote{Solar Oscillations Investigation/Michelson Doppler Imager (Scherrer et al. 1995)})  expanded the research opportunities.
 
With the passage to the new millennium, we attended a blooming of g-mode research based on the quality and accumulation of data. The first efforts were concentrated in looking for individual modes. Indeed, in 2000, Appourchaux et al. looked for individual spikes in the power spectrum of several instruments (VIRGO, BiSON and GONG) concluding that, with 90\% confidence level, no signals were detected.  Although they could not identify any g-mode signature, an upper limit of their amplitudes was established: at 200 $\mu$Hz, they would be below 10  ${\rm mm s^{-1}}$ in velocity, and below 0.5 parts per million in intensity. The same year, using the GOLF instrument and different power spectrum estimators, 
Garc\'\i a et al. (2001a, 2001b) found a peak around 285 $\mu$Hz that could be interpreted as one component of the $\ell=1, n=1$ mixed mode (up to 98\% confidence level). Later, in 2002, Gabriel et al. (2002), using a similar statistical approach, confirmed the existence of this mixed-mode candidate with a longer data set of the same instrument. Besides, two other structures at around 220 and 252 $\mu$Hz -- already studied with 2 years of GOLF data and reviewed by Gabriel et al. (1999) -- were highlighted to be potentially interesting due to their persistency with time. 

In order to reduce the threshold while maintaining the same confidence level, Turck-Chi\`eze et al. (2004) looked
 for multiplets instead of spikes and applied it to the g-mode research with GOLF. This study led to several patterns that were considered as g-mode candidates. Unfortunately, this analysis fails to properly label these modes with the correct degree and orders ($n$, $\ell$, $m$). In particular, the structure around 220 $\mu$Hz retained our attention. 
In fact, Cox and Guzik (2004) deduced theoretically that the g mode with the highest surface amplitude
 would be the $\ell=2, n=-3$ expected at 222.14  $\mu$Hz (from the Saclay seismic model computed by 
Couvidat, Turck-Chi\`eze and Kosovichev (2003)). Although the precise structure of this pattern changed with time, the main features still remained above the statistical threshold when 3000 days were used (Mathur et al. 2007 and references therein). 

In 2007(a), instead of looking for individual g modes, Garc\'\i a et al. found the nearly constant spacing in period of the dipole g modes between two consecutive modes of the same degree ($\ell$=1) and consecutive radial order $n$ using 9.5 years of GOLF data. 

Today, we face a situation where the fingertips of the gravity modes have been found collectively as well as some individual g-mode candidates. However it would be ex\-tre\-mely important if both discoveries could be confirmed by other instruments and also with longer data sets. Thus, in section 2 we extend the GOLF data series, on one hand, by more than 40\% since the last time we have studied the g-mode candidate around 220 $\mu$Hz and, on the other hand, by more than  20\% for the study of the asymptotic properties of the dipole modes. Then, in section 3, we study --using different instruments-- the region of the g-mode candidate $\ell=2, n=-3$ and the behavior with time of one of the multiplet components at 220.7 $\mu$Hz. We conclude in section 4 by doing some theoretical prospects on how the studies of the physics and dynamics of the solar interior could benefit in the near future by adding physical constraints coming from the detection of g modes.  

\section{GOLF 11.5 year-long series}
Nowadays, we have cumulated nearly 11.5 years (4182 d) of GOLF data from 1995 April 11 to 2007 September 22. These data correspond to time series calibrated into velocity following Garc\'\i a et al. (2005). As we have seen in the previous section, two different strategies have been followed to analyze the data: to look for individual modes above 150 $\mu$Hz, or to look for the asymptotic properties between 25 and 140 $\mu$Hz.

\subsection{Looking for individual multiplets: 220 $\mu$Hz region}

The most interesting case found in the previous analysis of the GOLF data --when individual multiplets were searched-- was the region around 220 $\mu$Hz. Indeed, an interesting pattern of peaks has been tracked with time since the very beginning of the mission when only two year-long data sets were available (Gar\-c\'\i a and Turck-Chi\'eze 1997; Pall\'e and Gar\-c\'\i \-a 1997). These analyses were based on the recurrence with time of some peak-patterns and their findings were summarized in Gabriel et al. (1999). 

Later, this region was studied with longer datasets. In particular with 1290 and 2975 days long (see Mathur et al. 2007 and references there in). First, with 1290 days and using the average of 4 tapered periodograms, a triplet was found centered at 220.72 $\mu$Hz and with two components at each side at:  220.10 and 221.28 $\mu$Hz. Another peak raised above the 98\% confidence level at 218.95 $\mu$Hz which, at that time, seemed quite far to belong to the same mode. However, all together could be detected as  a quadruplet with more than 98\% confidence level in a 10 $\mu$Hz window. Later, with 2975 days and using the average of 8 tapered periodograms, the same peaks stand above the statistical threshold and a new one (at 219.59 $\mu$Hz) raised between the left peak and the others forming a quintuplet as a whole.

\begin{figure}[htb*]
\includegraphics[angle=90,width=0.48\textwidth]{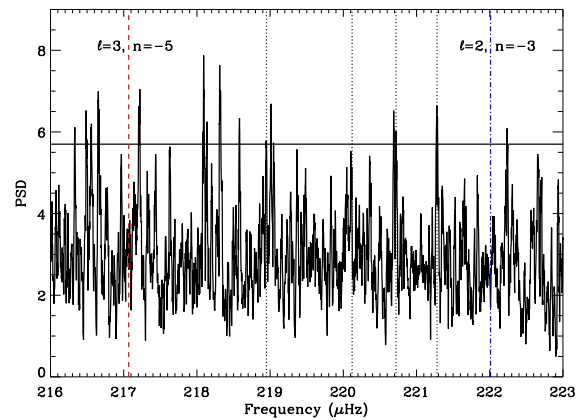} 
\caption{Power spectrum density of 4182 day-long GOLF velocity time series. The vertical dotted lines are the components of the multiplets previously seen with GOLF. The vertical dashed lines are the expected theoretical frequencies for the g modes  $\ell$=2 and 3 in this frequency region from the Saclay seismic model. The horizontal line is the 90\% probability threshold for a  quadruplet. }
\label{golf220}
\end{figure}

\begin{figure*}[ht*]
\includegraphics[angle=90,width=0.98\textwidth]{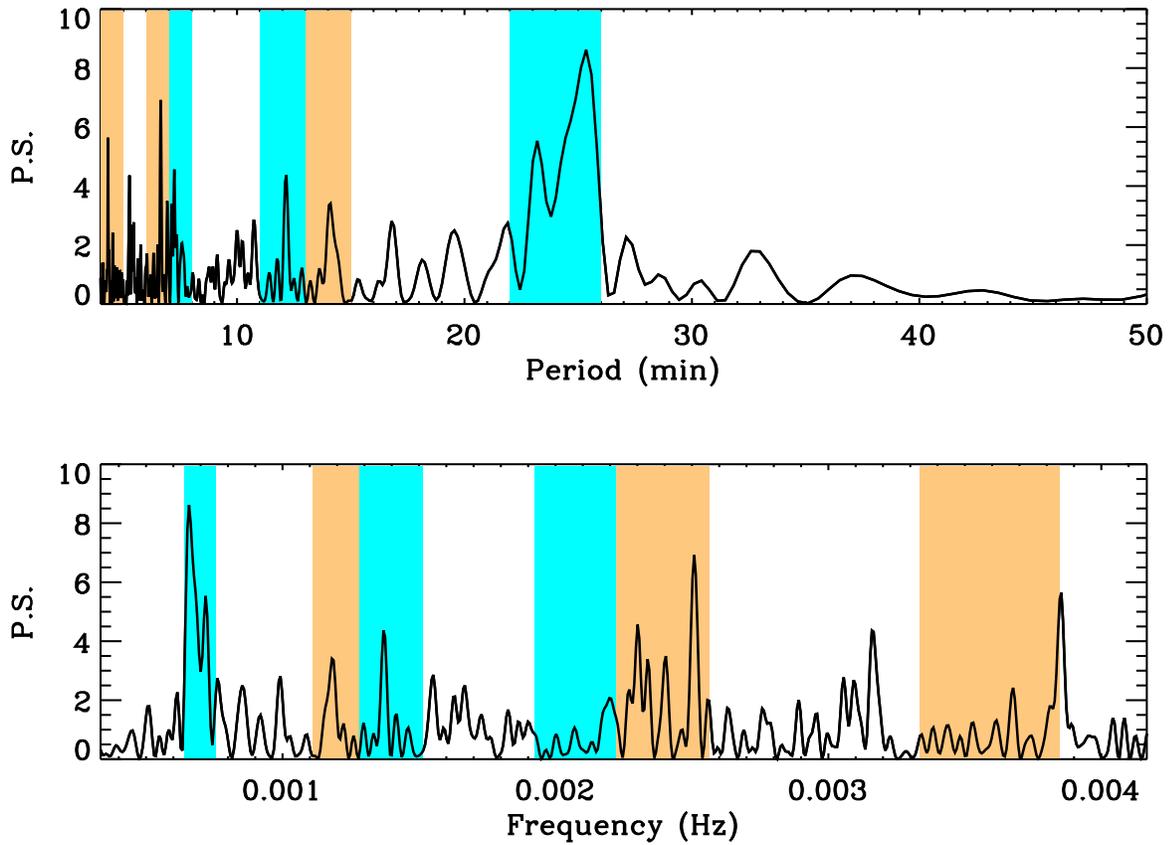} 
\caption{Power Spectrum of the GOLF power spectrum density computed between 25 and 140 $\mu$Hz. Top panel: as a function of the period; bottom panel: as a function of frequency. The shaded regions are the $\Delta P_1$ and $\Delta P_2, m=0$ regions and their first two harmonics (blue and orange respectively). }
\label{psgolf}
\end{figure*}

Today, the same type of periodogram estimator has been computed using 4182 day-long series (see Figure~\ref{golf220}). To keep the same significance in the statistical analysis we have maintained the same number of points in the studied windows. Therefore, due to the increase in the power spectrum resolution, we were obliged to reduce the frequency window from 10 to 7.1 $\mu$Hz. The resultant spectrum is closer to the situation we had with 1290 days compared to the one we had with 2975 days series. In fact, the peak that raised in between the two groups (at 219.59 $\mu$Hz) is now embedded  into noise and the one at 220.10 $\mu$Hz is at the detection limit. Therefore a quadruplet is still detectable if we consider the uncertainty related to the computation of the 90\% threshold (see the discussion on Turck-Chi\`eze et al. 2004). 
The theoretical frequencies for the g modes $\ell$ =2, $n$ = -3 and $\ell$ = 3, $n$ = -5, calculated with the seismic model are superimposed on Figure~\ref{golf220}. A study on the uncertainty on these predictions has shown that depending on the physics included in the solar models (abundances, microscopic diffusion, horizontal diffusion in the tachocline), we can have an uncertainty on the frequencies of 3-4 $\mu$Hz in this frequency range (Mathur et al. 2007).

\subsection{The dipole asymptotic properties}
Using 3481 days of GOLF time series, Garc\'\i a et al. (2007) uncovered the presence of a pattern in the power spectrum (P.S.) of the power spectrum density between 25 and 140 $\mu$Hz that could be interpreted as the quasi constant separation in period between consecutive radial orders $n$  of dipole ($\ell$=1) modes. This separation ($\Delta P_1$) was characterized as a pattern of peaks with a maximum of 6.5 $\sigma$ and a total power of 2.9 times more than the average power of the rest of the power spectrum. $\sigma$ is the standard deviation defined as $(1/(N-1) \sum_{j=0}^{N-1} (x_j-x)^2 )^{1/2}$, $N$ is the number of points between 4 and 50 min in the periodogram, and $x_j$ are the individual bins.

With the new series 20\% longer (see Figure~\ref{psgolf}), the structure has grown up to a maximum power of 8.6 $\sigma$ and a total power of 4.1 times higher than the average power of the P.S. It is also important to note that the highest peak of the spectrum between 4 and 50 minutes is the one attributed to the $\Delta P_1$ structure. Besides, the other highest peaks correspond to the regions where we expect power from the harmonics of this peak as well as the ones corresponding to the periodicity, $\Delta P_2$ of the quadrupole modes ($\ell$=2). In Figure~\ref{psgolf}, we have only shaded the regions where the $\Delta P_2$ for the $m$=0 components and its harmonics are expected: they are the only $m$ components, the zonal ones, that are not modified by the rotational splitting. Indeed we see near the region of the second harmonic two high peaks at $\sim$ 0.00315 Hz and $\sim$ 0.00385 Hz that could be the signature of the $m=\pm 2$ components. A deeper study will be done in the future to explore the structure of the $\Delta P_2$ periodicity and its harmonics. It is important to remember that the position of the shaded regions are almost insensitive to the choice of the solar model (Mathur et al. 2007). However, their sizes depend rather on the simulated rotation rate, mainly for the $\Delta P_2$ and this is the reason why we have only considered the zonal components. In the case of the $\Delta P_1$ the resolution of the P.S. is not enough to correctly separate the components even for very high core rotation rates. The two-peak structure observed in the region between 22 and 26 minutes could be due to the high level of noise rather than the signature of the $m \pm 1$ components. Further studies would also be necessary before any claim in this direction. 

The conclusions obtained after the analysis of the reconstructed waves built after filtering the $\Delta P_1$ region between 22 and 26 minutes and its first harmonic are qualitatively the same as in the previous shorter study (Garc\'\i a et al. 2007).

\section{The region around 220 $\mu$Hz.}

\begin{figure}[htb*]
\includegraphics[angle=90,width=0.48\textwidth]{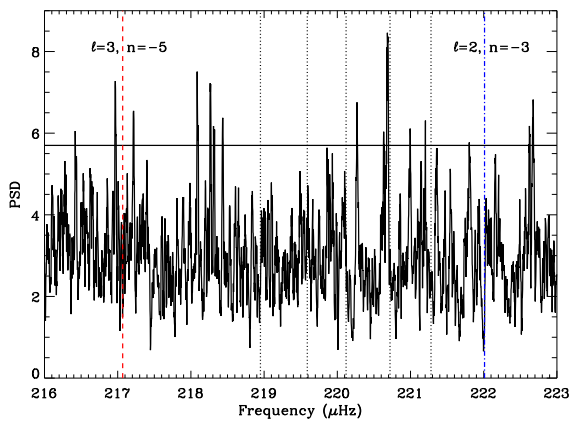} 
\caption{Power spectrum density of 4098 days of intensity time series computed from the SPM blue channel. The vertical dotted lines are the components of the multiplets previously seen with GOLF. The vertical dashed lines are the expected theoretical frequencies for the g modes  $\ell$=2 and 3 in this frequency region from the Saclay seismic model. The horizontal line is the 90\% probability threshold for a  quadruplet. }
\label{spmbl220}
\end{figure}

In the previous section we have seen that there is a statistically significant (above 90\% confidence level) structure around 220 $\mu$Hz using GOLF data and this structure --which evolves with time-- has been present in these data continuously during the last 11.5 years. In adition, when using the longest data set available for the SPM of the VIRGO experiment (4098 day), this feature also appears and with higher significance than in GOLF.  So, in this section both data sets will be used to analyze the time evolution details of this feature. Figure~\ref{spmbl220} shows the power spectrum density in the same frequency region than in the GOLF case (see Figure~\ref{golf220}). It corresponds to 4098 day-long time series of the blue SPM channel from 1996 April 11 until 2007 June 30. Compared to the multiplet components detected with GOLF, only the one at 220.7 $\mu$Hz component is clearly visible with a high signal-to-noise ratio ($>$ 8 $\sigma$). Some power excess is also present at the most right peak at 221.28 $\mu$Hz but slightly shifted compared with the frequency extracted from GOLF.

Besides those components, some other peaks are pre\-sent in both spectra. In particular, around 218.2 and 217 $\mu$Hz, there are some patterns in common in both instruments with peaks in a range of 6 to 8 $\sigma$. Therefore, this frequency region is of particular interest, deserving the use of all the best available data sets.
  
\subsection{Time evolution as seen by VIRGO instruments}

The  VIRGO experiment and its performance has been already reported in detail (Fr\"{o}hlich et al. 1995; Fr\"{o}hlich et al. 1997).  One of its instruments is a three channel full-Sun PhotoMeter (SPM) which measures solar irradiance through 5 nm wide filters centred at 402 nm (blue), 500
nm (green), and 862 nm (red) that look at the Sun as a star with a 60 s cadence. VIRGO experiment also contains two different ac\-ti\-ve-cavity radiometers (DIARAD and PMO)  for monitoring the solar ``constant'' and a Luminosity Oscillation Imager (LOI) with low resolution (12 pixels) for the measurements of the radiance distribution over the solar disk at 500 nm.

From a  VIRGO/SPM time series of 3620 days (from 1996 April 11 to 2006 March 3), 57 time series of 800 days, each one shifted by 50 days consecutively,  have been extracted and the respective  power spectra computed. In this way, the temporal evolution of the spectral signals can be studied. The power spectra of the 800 days time series are computed with a zero padding factor of 5, thus increasing the number of bins by the same factor. The resultant frequency resolution is 0.00289  $\mu$Hz but the points are correlated and not statistically independent (see the discussion in Gabriel et al. (2002) about the use of zero padded spectra).

\begin{figure}[htb*]
\includegraphics[angle=90,width=0.48\textwidth]{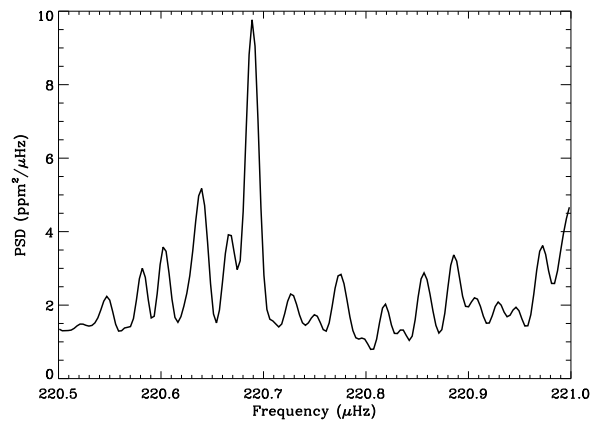} 
\caption{Average of 57 power spectra corresponding to the blue channel of VIRGO/SPM (figure~\ref{x1})  where the signal at 220.7 $\mu$Hz is clearly visible.}
\label{x2}
\end{figure}

\begin{figure*}[p*]
  \vspace{+0.02\textwidth}   

\includegraphics[width=0.90\textwidth]{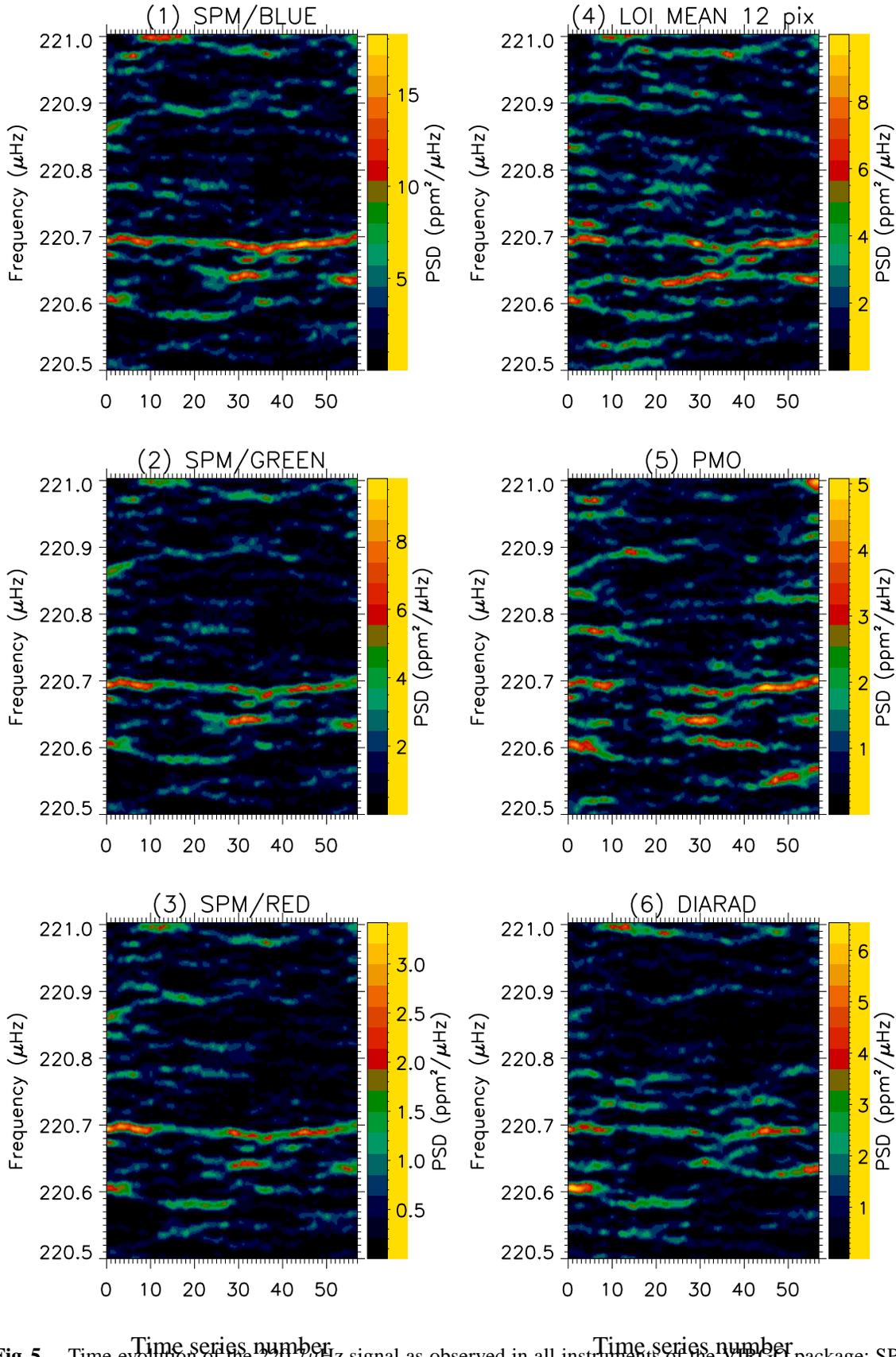} 
\caption{Time evolution of the 220.7$\mu$Hz signal as observed in all instruments of the VIRGO package:  SPM blue, green and red; LOI, PMO and DIARAD instruments. Each vertical column corresponds to the PSD of 800 days shifted by 50 days.}
  \vspace{-0.08\textwidth}   
  
  \centerline{ \large    
   \hspace{0.13\textwidth}{Time series number}                       
   \hspace{0.25\textwidth}{Time series number}
\hfill  }
  \vspace{0.28\textwidth}    

\label{x1}
\end{figure*}

In the VIRGO/SPM data a clear and stable signal at 220.7 $\mu$Hz is observed during the 10 years of the SoHO mission. The signal is well observed in all three channels of SPM (blue, green and red) as  shown in the Figures~\ref{x1} (1, 2 and 3).  The corresponding average power spectrum of the 57 power spectra has also been computed for each channel and a sample of them is shown in Figure~\ref{x2}. In order to look if this signal is also observed in the others VIRGO instruments, the same analysis have been undertaken using the other data instruments:  LOI imager (averaging the 12 pixels) and  the PMO and DIARAD active radiometers, providing positive results on the presence of the same feature in all of them (see Figures~\ref{x1} (4, 5 and 6)). 

The possibility that this signal could be induced by some instrumental effects has also been investigated. In particular, temperature fluctuations of the detector. Each of the three SPM channels measurements are corrected by a quantity proportional to their respective value of the detector temperature. This correction is applied at the level 1 stage of the data pipeline, so  data used in the present analysis are already multiplied for this quantity. This correction is:
 \begin{eqnarray}
SPM_{channel}=(1+C_{channel}(TS_{channel}-293.15)), \nonumber
\end{eqnarray}
 
\noindent where channel means blue, green or red; $C_{channel}$ is a  constant for each channel and $TS_{channel}$ is the temperature of each of the three detectors.

In Figure~\ref{x3}  we show the SPM blue signal (1), the SPM blue detector temperature fluctuations (2)  and in (3) a zoom of the fainter region of (2) (time series 20 to 57). We do not see any correlation between signal and temperature being the temperature fluctuations almost three order of magnitude smaller than the 220.7 $\mu$Hz signal.

\subsection{Time evolution as seen by velocity instruments}

\begin{figure*}[p*]
  \vspace{-0.02\textwidth}   
\includegraphics[width=0.90\textwidth]{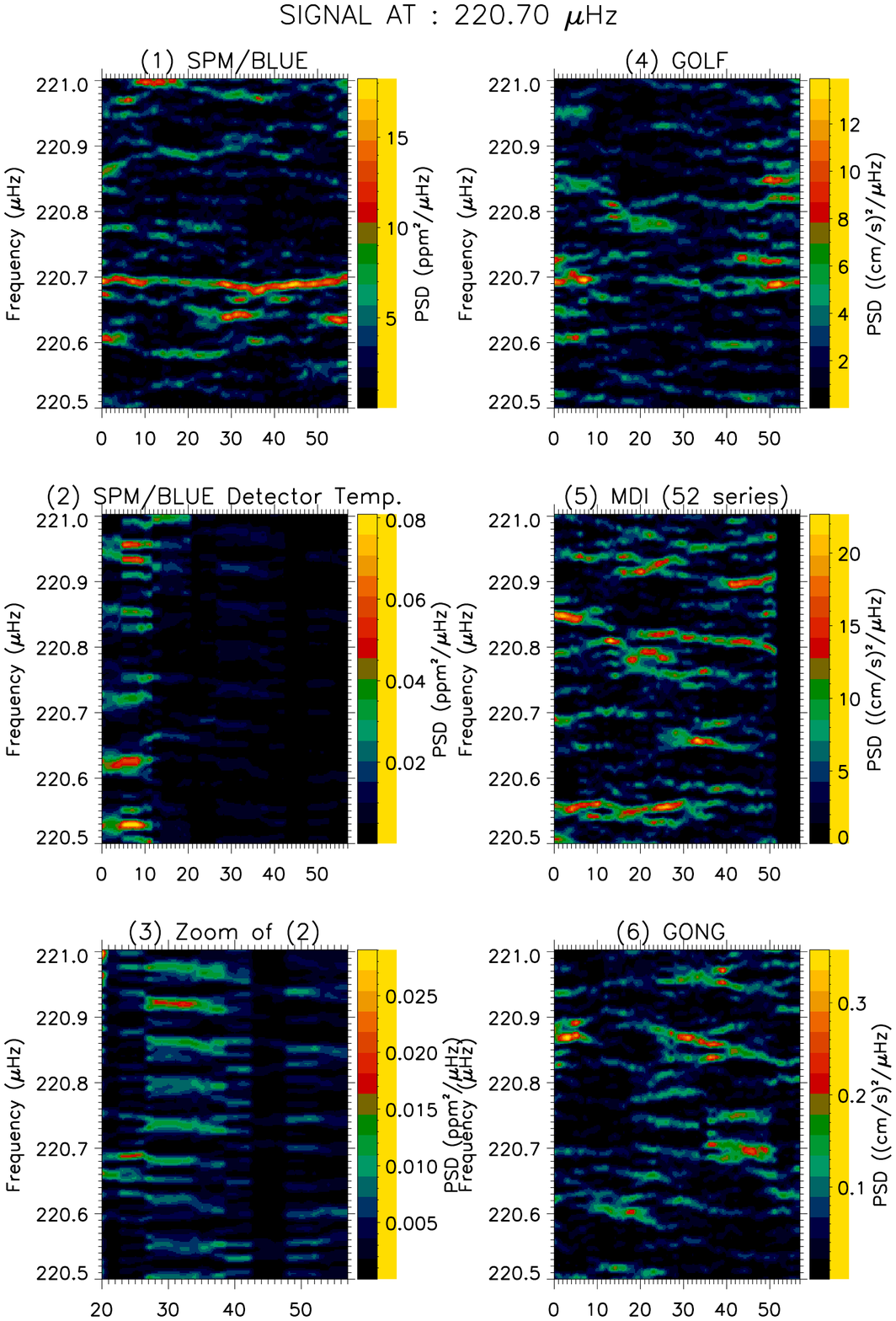} 
  
  \centerline{ \large    
   \hspace{0.13\textwidth}{Time series number}                       
   \hspace{0.25\textwidth}{Time series number}
\hfill  }
\caption{Time evolution of the 220.7$\mu$Hz signal for SPM/blue (1); detector  temperature  fluctuations for SPM/blue (2); (3) is a zoom of (2). In (4), (5) and (6) the same analysis but for velocity data from GOLF, MDI and GONG instruments. Although weaker in velocity than in intensity, a trace is observed mainly in GOLF and  GONG data. Each vertical column corresponds to the PSD of 800 days shifted by 50 days.}
 \label{x3}
\end{figure*}

 This signal at 220.7 $\mu$Hz has a high visibility in the intensity data of the VIRGO experiment. It is also interesting to verify the behavior of this region in velocity. A preliminary analysis of GOLF, MDI (integrated data) and GONG\footnote{Global Oscillation Network Group (Harvey et al., 1996)} (sun-as-a-star data) --ground based Network-- has also been done in the same way as for VIRGO. Figure~\ref{x3} shows the resultd for GOLF (4), MDI (5) and GONG (6). The visibility of this signal in velocity seems to be much lower than in intensity but a trace of the signal is observed mainly in GOLF and in GONG data.

\section{Inferring properties of the radiative interior}
During the last years we have seen the detection of some patterns that have been considered as g-mode candidates. In this paper we have seen that several instruments have interesting features in the power spectrum. Some of them have been presented in the spectrum for the last 4000 days and the careful analyses of the VIRGO-SPM house-keeping parameters did not show any similar instrumental behavior. Everything seems to point to a solar origin. Whether it is signal from the g modes or just a very unlikely combination of granulation noise, the future will tell us. The possibility of measuring individual g modes could be just in front of us. It seems that we have likely measured power coming from the cumulative effect of several of such modes. Therefore, we should be prepared to incorporate all this information to better constrain the structure and the dynamics inside the radiative zone and, in particular, in the core.

\subsection{Constraints on the rotation profile}
The rotation of the solar interior is well known down to the frontier the solar core (e.g. see Thompson et al. 2003 and references therein). Unfortunately, p modes do not contain enough information to properly constrain the rotation inside the nuclear core (e.g. see Couvidat et al. 2003; Garc\'\i a et al. 2004), even using high radial order, low-degree p modes (Gar\-c\'\i a et al., 2008). The analysis of the $\Delta P_1$ structure and the comparison with solar models (with different artificial solar rotation profiles inside the core) favored --in average-- a faster rotation rate in the deepest layers. However it was impossible to constrain with accuracy the profile.    

To progress in our knowledge of the inner core we need to include information coming from the analysis of low-degree low-order p modes, mixed and individual g modes (e.g. see Pro\-vost, Ber\-thomieu and Morel, 2000). Recently, Mathur et al. (2008) have shown the influence on the inferred rotation rate of the solar core when 1 and 8 simulated g modes are introduced in the 2D inversions based on the Regularized Least-Squares method (for more details see Eff-Darwich and P\'erez Hern\'andez, 1997). 

\begin{figure}[htb*]
\includegraphics[angle=90,width=0.48\textwidth]{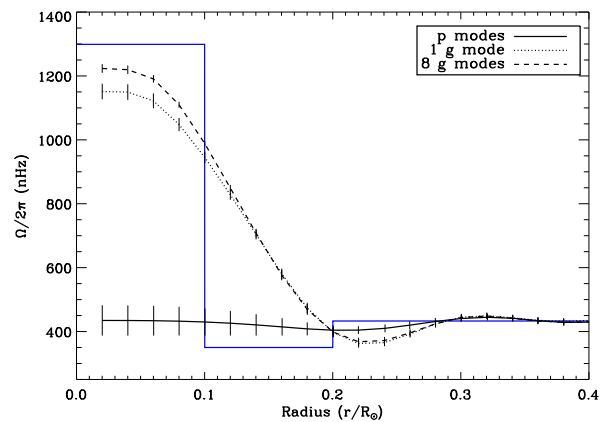} 
\caption{Equatorial rotation rate below 0.4 $R_\odot$ inferred from the use of p modes only, when 1 g mode is introduced and when 8 g modes are introduced. The simulated rotation profile is the continuous blue line. }
\label{inver}
\end{figure}

Figure~\ref{inver} summarizes the conclusions of this work. Indeed, an artificial profile has been computed in the core: a step profile having a rate of 350 nHz in the 0.1-0.2 $R_\odot$ region and a rate 3 times larger than the rest of the radiative zone below 0.1 $R_\odot$. Although this profile is unlikely to
be realistic, it enables us to characterize the quality of the inversions, as the code has difficulty in reproducing these steep gradients. Using this profile, the rotational splittings of the p and g modes have been computed and the inversion calculated with three different sets of modes. The first set contains only the same p modes than those available in the real case. In fact, the error bars have been taken from real observations. This is the reference inversion (continuous line with errors in Figure~\ref{inver}). However this result is reliable down to 0.16 or 0.2 $R_\odot$ depending on the set of p mode used (See also the discussion in Garc\'\i a et al. 2008). The rotation rate inferred below this radius is an a priori knowledge introduced via the regularization term. Then, 1 g mode --the candidate $\ell$=2 and $n$=-3-- has been added with a typic error bar of 7.5 nHz (dotted line in Figure~\ref{inver}). In this case a constrain in the core is fixed by this mode and a better rate inside the core is obtained. When 8 g modes ($\ell$=1 and 2 with $n$ ranging form  -2 to -6)  are added (all with 7.5 nHz error bars) the agreement is even better (dashed line in Figure~\ref{inver}) than in the previous case with smaller error bars. The profile is still poorly determined and a higher number of modes or smaller error bars are needed to improve the information on the profile. Nevertheless, the resolution kernels in the latest case are much better determined as compared to the case of adding one single g mode. In this sense, for the deepest kernels, there are less wiggles and the secondary lobes become lower than with only 1 g mode (for more details see Mathur et al. 2008). For the deeper layers, below 0.12$R_{\odot}$, by adding more g modes, the secondary lobes are decreased by a factor of two and some of them even disappear. While the upper limit of the main lobe goes up to $\sim$0.25$R_{\odot}$ in the first case, it decreases down to $\sim$0.15$R_{\odot}$ in the second one. Thus, we can say that these resolution kernels are better localised.

\subsection{Constraints for the solar structure}
The situation for the case of the structural physical parameters is much better than for the solar rotation profile. In this case, for example, the profiles of sound speed and density are rather well determined inside the solar interior even inside the solar core (e.g. Basu et al 1997; Couvidat, Turck-Chi\`eze and Kosovichev 2003; Mathur et al. 2007) thanks to the information carried out by the $\ell$=0 modes that go through the deepest layers of the Sun. However, it would be better if we could add some constraints coming mainly from the core because the p modes --due to the increasing sound speed towards the interior-- spend a much smaller fraction of time in these central regions than in the external layers. 

While we wait for the advent of the first individual g-mode frequencies, we can already check if recent measurements of the $\Delta P_1$ separation could be used to distinguish between models containing different physical pro\-ce\-sses inside. Theoretical works have shown that the most recent solar models have $\Delta P_1$ differences of less than a minute (Mathur et al. 2007; Zaatri et al. 2007) which are too small to be directly detectable. However, instead of using the separations themselves, we could utilize the correlation of the reconstructed waves (Garc\'\i a et al. 2007; Garc\'\i a, Mathur \& Ballot 2008). By doing so, we have more sensitivity to the models and we could start to constrain some physical processes in the core of the Sun.

\begin{figure}[htb*]
\includegraphics[width=0.48\textwidth]{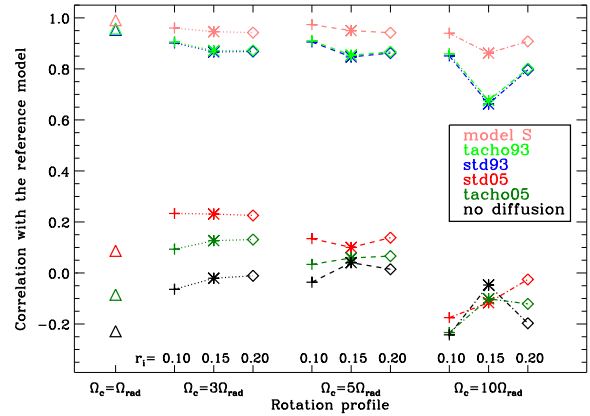} 
\caption{Correlation rates between the reference model (seismic) and another model with the same rotation profile. These models, described in Mathur et al. (2007), are: {\it std93} (blue), {\it std05} (red), {\it tacho93} (light green), {\it tacho05} (dark green), {\it no diffusion} (black) and {\it model S} (pink). The rotation profiles used have a differential rotation in the convective zone and a rigid rotation from 0.7 down to $r_i$ (= $r/R_\odot$): 0.1 (crosses), 0.15 (asterisks) and 0.2 (diamonds). The rotation rate below $r_i$ is constant: $3 \Omega_\mathrm{rad}$ (dotted line), $5 \Omega_\mathrm{rad}$ (dashed line) and $10 \Omega_\mathrm{rad}$ (dotted-dashed line), where $\Omega_\mathrm{rad}$ is the rotation rate in the rest of the radiative zone (433 nHz). Figure taken from Garc\'\i a, Mathur and Ballot (2008). }
\label{struc}
\end{figure}
  
The main uncertainty will come from the dynamics as the position of the $m$-components of each mode will depend on the rotational splitting. For this reason, to test the method, the correlations between a reference model (the saclay seismic model (Couvidat, Turck-Chi\`eze and Ko\-so\-vi\-chev 2003)) and different rotation rates (the same used in Garc\'\i a et al. (2007)) were calculated. Figure~\ref{struc} shows the results of the comparison with 6 solar models including the model S (Christensen-Dalsgaard et al. 1996), standard models (na\-med \emph{std}) and others that mainly differ on three physical inputs: the microscopic diffusion (all including this process but the one named \emph{no diffusion}), the turbulent horizontal diffusion in the tachocline (named \emph{tacho}), and the abundances of chemical elements. The models using the old abundances (Grevesse and Noels 1993, named \emph{93}) are clearly separated from those using newer ones (Asplund, Gre\-ve\-sse and Sauval 2005, na\-med \emph{05}) and the obsolete model without including the microscopic diffusion. 
This result is in agreement with previous analyses looking directly at the p-mode frequency differences computed for different models. 
However, in our analysis it is impossible to distinguish between models using very close phy\-sics like the standard models or those including the turbulence treatment in the tachocline. 


\section{Conclusions}
In the present work we have shown that the longer GOLF time series confirm the results concerning the g-mode research. On one hand, there are still interesting features in the GOLF spectrum around 220.7 $\mu$Hz that are at the detection limit of a quadruplet above the 90$\%$ confidence level. On the other hand, the amplitude of the  pattern attributed to the signature of the dipole gravity modes has increased and it contains more relative power than before. The highest peaks on the periodogram of the power spectrum could be interpreted as coming from $\Delta P_1$, $\Delta P_2$ and their harmonics.

The study of the VIRGO data have revealed the presence of one of the components of the quadruplet detected by GOLF in the same frequency bin. The analysis of the temporal evolution of this signal in all the experiments of VIRGO has shown that this signal has been above noise level since the beginning of the mission and during the last 10 years. The SPM data has the highest signal-to noise ratio which is 3 orders of magnitude higher than the variations found in the same region for the house-keeping data, in particular, the temperature of the detector.

The same analysis performed on velocity experiments has shown a similar behavior but with smaller signal-to-noise ratio. It is important to notice that the GONG data is a gro\-und-ba\-sed facility completely independent to SoHO and it would be very important if this preliminary result can be confirmed.

Finally, we have illustrated how  2D inversions of the equatorial rotation profile can be improved by including the information given by 1 or several g modes and also how the use of the asymptotic properties of the g modes could be useful to put new constraints to the solar models.

\acknowledgements

Some of the ideas presented here found their origin in the fruitful discussions of the Phoebus collaboration in the framework of the International Space Science Institute\footnote{http://www.issi.unibe.ch/teams/GModes/} whose support is greatly acknowledge. The authors want to thank D. Salabert for providing them the GONG integrated data.
The GOLF experiment is based upon a consortium of institutes (IAS, CEA/Sa\-clay, Nice and Bordeaux Observatories from France, and IAC from Spain) involving a large number of scientists and engineers, as enumerated in Gabriel et al. (1995). SoHO is a mission of international cooperation between ESA and NASA. This work has been partially funded by the grant AYA2004-04462 of the Spanish Ministry of Education and Culture and by the European Helio- and Asteroseismology Network (HELAS), a major international collaboration funded by the European Commission's Sixth Framework Programme.




\end{document}